\documentclass[twocolumn]{aastex62}

\graphicspath{{./}{figures/}}

\received{TBD}
\revised{TBD}
\accepted{TBD}
\submitjournal{ApJ}

\shorttitle{Stellar contamination with JWST}
\shortauthors{Iyer $\&$ Line}

\begin{document}

\title{THE INFLUENCE OF STELLAR CONTAMINATION ON THE INTERPRETATION OF NEAR INFRARED TRANSMISSION SPECTRA OF SUB-NEPTUNE WORLDS AROUND M-DWARFS}

\correspondingauthor{Aishwarya Iyer}
\email{aiyer13@asu.edu}

\author{Aishwarya R. Iyer}
\affil{School of Earth and Space Exploration \\
Arizona State University \\
525 E. University Dr., Tempe AZ 85281}

\author{Michael R. Line}
\affil{School of Earth and Space Exploration \\
Arizona State University \\
525 E. University Dr., Tempe AZ 85281}



\begin{abstract}

The impact of unocculted stellar surface heterogeneities in the form of cool spots and hot faculae on the spectrum of a transiting planet has been a daunting problem for the characterization of exoplanet atmospheres. The wavelength-dependent nature of stellar surface heterogeneities imprinting their signatures on planetary transmission spectra are of concern particularly for systems of sub-Neptunes orbiting M-dwarfs.  Here we present a systematic exploration of the impact of this spot-contamination on simulated near infrared transmission spectra of sub-Neptune planets. From our analysis, we find that improper correction of stellar surface heterogeneities on transmission spectra can lead to significant bias when inferring planetary atmospheric properties. However, this bias is negligible for lower fractions of heterogeneities ( $<1\%$). Additionally, we find that acquiring $\textit{a priori}$ knowledge of stellar heterogeneities does not improve precision in constraining planetary parameters if the heterogeneities are appropriately marginalized within a retrieval---however these are conditional on our confidence of stellar atmospheric models being accurate representations of the true photosphere. In sum, to acquire unbiased constraints when characterizing planetary atmospheres with the James Webb Space Telescope, we recommend performing joint retrievals of both the disk-integrated spectrum of the star and the stellar contamination corrected transmission spectrum.

\end{abstract}



\section{Introduction} \label{sec:intro}
 Transit spectroscopy of planets around small, cool stars offer a promising opportunity to characterize small cool worlds \citep[e.g.,][]{seasas2000,swain2008,huitson2012,gillon2016,gillon2017,luger2017}. An important consideration in characterizing such planets is the nature of the host star \citep{apai2018}. Transmission spectroscopy of hot-Jupiters to terrestrial worlds alike have 
demonstrated evidence of stellar activity \citep{pont2007,alonso2008, zhang2018, rackham2017}. M-dwarf hosts in particular, exhibit higher recorded detections of stellar activity in the form of flares \citep{hawley1991,schmidt2014,hawley2014} and H$_\alpha$ chromospheric emissions \citep{delfosse1998} relative to F, G, K-type host stars, indicative of frequent occurrence of photospheric heterogeneities such as stellar spots and faculae. Study of M-dwarf systems are extremely timely especially with the recent launch of NASA's Transiting Exoplanet Survey Satellite (TESS), which will discover thousands of exoplanets around such stars \citep{ muirhead2018,ballard2019predicted} providing the need for follow up with future missions like JWST.

Stellar spots that are occulted by a transiting planet contribute to observable changes in the transit light curve. Techniques to model the occulting heterogeneity have been successful to some degree by either excluding the spot crossing event from the transit fit \citep{pont2008,carter2011,narita2013} or by correcting for epoch-to-epoch variations in the star from relative offsets between observations therefore accounting for any differences in the planetary spectrum
\citep{zellem2017}. These techniques however, do not account for homogeneously scattered regions of heterogeneities that are persistent with the entire data in all observations.

Unocculted stellar spots as well as homogeneously covered regions of smaller spots affect the transit baseline in addition to causing discrepancies in the transit depth measurements depending on the fractional coverage in the area of the spots and their temperature contrast relative to the photosphere \citep{pont2008,silva2008,czesla2009,wolter2009,agol2010,berta2011,carter2011,desert2011,sing2011,fraine2014,mccullough2014,oshagh2014,barstow2015,damasso2015,zellem2015,rackham2017}.
Attempts to understand the nature of unocculted spots have thus far been done with photometric monitoring of the star, i.e., tracking the rotational variability in the modulating brightness on the stellar photosphere translating onto the light curve \citep{pont2008,berta2011,desert2011,sing2011,knutson2012,pont2013,zellem2015}. Despite reasonable success, efforts with photometric monitoring of unocculted stellar spots have been inadequate in teasing out underlying heterogeneities spread across the entire disk of the star, especially those that are not significant enough to show variability effects on the stellar light curve \citep{jackson2012,rackham2017,rackham2018}. Studies by \citealt{chapman1987} and \citealt{shapiro2014}, have also shown that every differential region of the star's photosphere--whether the transit chord or the entire disk has its own unique spectrum. Given these challenges, the effect of unocculted stellar spots remains a persistent problem despite corrections from photometric variability monitoring, which at best provides only a lower limit estimate for regions of heterogeneities on the photosphere \citep{rackham2018}.

Several recent studies \citep{pont2013,mccullough2014,oshagh2014,barstow2015,barstow2015erratum,rackham2017,barstow2018,bruno2018,murgas2019,pinhas2018,rackham2018,zhang2018} have demonstrated that unocculted as well as homogeneous spots and faculae on the host star present themselves in a wavelength-dependent manner in the planetary transit data. The wavelength-dependence of stellar heterogeneity signatures is particularly influential for transit spectroscopy where these features can often be degenerate with planetary atmosphere features. This necessitates accounting for stellar heterogeneities in order to probe the true nature of a planetary atmosphere.
Improper characterization of the effects of stellar surface heterogeneities on the transmission spectrum limits our ability to place precision constraints on basic planetary conditions like temperature structure, composition, and cloud properties.  Therefore, it is vital to thoroughly understand influence of stellar surface heterogeneities on our ability to infer basic planetary properties in the era of high fidelity observations anticipated from the James Webb Space Telescope (\textit{JWST}).


To understand these degeneracies and the impact of {\it not} accounting for unocculted star spots on atmospheric inference, we leverage the utility of atmospheric retrieval.    In section \S\ref{sec:analytical}, we briefly explore the effect of stellar heterogeneities on the shape of a planetary transmission spectrum analytically, providing intuition for degeneracies that might occur in interpretations of transit spectra followed by a description of our atmospheric retrieval model. In \S\ref{sec:retrieval} we investigate the level of bias that arises while inferring planetary properties from JWST transmission spectra when ignoring corrections for stellar heterogeneities. Additionally, we address the question of whether \textit{a priori} knowledge of stellar heterogeneity, given the current state of M-dwarf stellar models, could improve precision on the inferred planetary characteristics from \textit{JWST} transmission spectra.  In this section, we also investigate the potential biases arising directly from the stellar models when inferring planetary properties. Section \S\ref{sec:discandconc} presents a discussion on various sources of bias that can affect appropriate characterization of the transmission spectrum in addition to the stellar contamination correction, followed by a brief summary of our findings. In sum, our work demonstrates that regardless of the coverage area or temperature contrast of a given stellar heterogeneity feature, one can infer unbiased planetary properties retrieved from contaminated transmission spectra provided an appropriate correction for stellar heterogeneity has been incorporated--however these are conditional on our confidence of stellar atmospheric models being accurate representations of the true photosphere.

\section{ANTICIPATED DEGENERACIES} \label{sec:analytical}

Here we briefly build our intuition for how the effects of stellar contamination can alter the shape of the planetary transmission spectrum. The transmission spectrum corrected for stellar surface heterogeneities is given by:
\begin{eqnarray}\label{transmission}
\alpha_{\lambda,c} =\epsilon_\lambda \alpha_\lambda
\end{eqnarray}
where $\alpha_\lambda$ is the native planetary transit spectrum and $\epsilon_\lambda$ is the contamination factor ( \citealt{mccullough2014}, \citealt{zellem2017}, and \citealt{rackham2018}, see their Equation 1 and 2) given by:
\begin{equation}\label{contam_fac}
\epsilon_{\lambda} = \frac{1}{\Big(1 - f_{s}\Big(1 - Q_{\lambda,s}\Big)\Big)}
\end{equation}

where, Q$_{\lambda,s}$ is the spot-to-photospheric stellar flux ratio ($F_{\lambda,s} / F_{\lambda,p}$) and $f_{s}$ is the fractional coverage of spots. 
The shape of the  transmission spectrum  (when corrected for stellar heterogeneities) is given by the derivative of equation \ref{transmission} combined with the expressions given in \citealt{LineParm2016} (see their equations 6-9, for gas+gray cloud system): 


\begin{equation}\label{shape}
\frac{d\alpha_{\lambda,c}}{d\lambda} =\epsilon_\lambda \frac{2R_p}{R^2_*} H \Big( \frac{1}{1+\gamma_{\lambda}} \Big) \frac{d ln(\sigma_\lambda)}{d\lambda} + \frac{d\epsilon_\lambda}{d\lambda} \Big(\frac{Rp^2}{R_*^2}+\frac{2R_pz_\lambda}{R_*^2}\Big)
\end{equation}

where R$_p$ and R$_*$ are the radius of the planet and star respectively, z$_\lambda$ is the wavelength dependent sharp occulting disk radius, $H$ is the planetary atmospheric scale height, $\sigma_\lambda$ is the absorption cross-section for a given gas, and $\gamma_{\lambda}$ is the gas-to-gray cloud opacity ratio. 

The shape of the stellar contamination spectrum, which is simply the wavelength-dependent derivative of the contamination factor (equation \ref{contam_fac}) is given by:
\begin{equation}\label{epsilonderivative}
    \frac{d\epsilon_\lambda}{d\lambda} = \frac{\Big(f_{s}\frac{dQ_{\lambda,s}}{d\lambda}\Big)}{{(1-f_{s}+f_{s}Q_{\lambda,s})^2}}
\end{equation}






 \begin{figure}[!tbp]
    \includegraphics[width=\columnwidth]{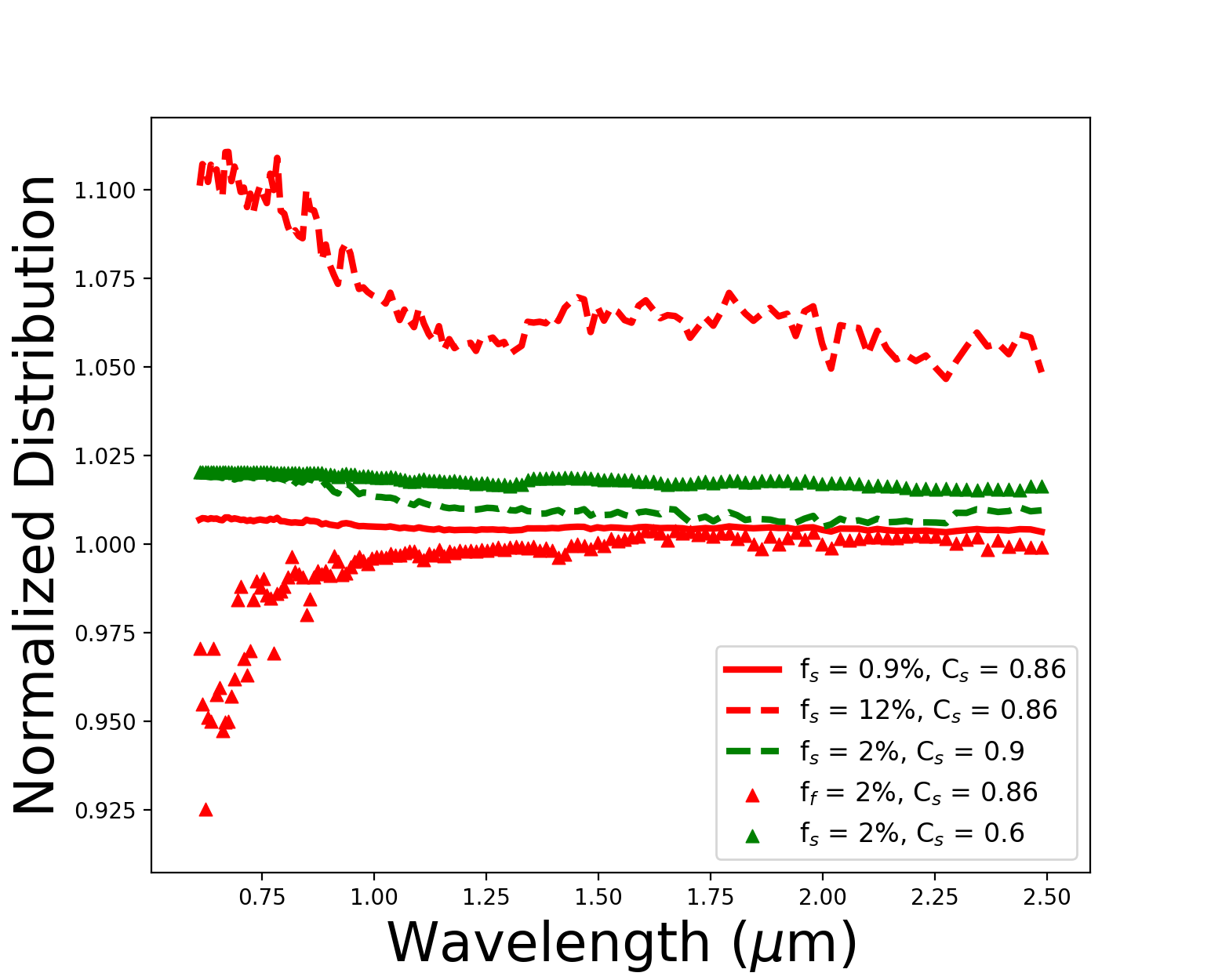}
    \caption{Stellar contamination spectra as a function of wavelength modeled for an (using PHOENIX models from \citealt{allard2003,allard2007,allard2009}) M-dwarf with photosphere temperature T$_p$ = 3300 K with varying cases of surface heterogeneities (following equation \ref{contam_fac}). The contamination spectra in red are for varying fractional coverage (f$_s$) in the heterogeneity area on the photosphere at a fixed temperature contrast (C$_s$ = T$_s$/T$_p$ of 0.86). The contamination spectra in green are for changing spot temperature contrast at a constant f$_s$ (2$\%$). Small values of f$_s$ produce a contamination spectrum close to 1 (e.g., no contamination), whereas a large f$_s$ results in an power-law like effect at shorter wavelengths. Faculae cause a negative power-law like behavior at shorter wavelengths. For a fixed f$_s$, increasing the temperature contrast alone causes an increased stretching effect at longer wavelengths.} 
    \label{contamplot}
  \end{figure} 

Simplistically assuming that the stellar fluxes behave like blackbodies and also that the temperature contrast between the photosphere and spot is small, consistent with Doppler Imaging observations for M-dwarfs \citep{strassmeier2010zoo,andersen2015stellar} i.e. T$_p$ $\approx$ T$_s$, and therefore we can write:
\begin{equation}
     Q_{\lambda,s} \approx 1+\frac{hc}{\lambda k_B} \Big(\frac{1}{T_p} - \frac{1}{T_s}\Big)
\end{equation}
\begin{equation}\label{qderivativeapprox}
     \frac{dQ_{\lambda,s}}{d\lambda} \approx - \frac{hc}{\lambda^2k_B} \Big(\frac{1}{T_p} - \frac{1}{T_s}\Big)
\end{equation}

At longer wavelengths $dQ_{\lambda, s}/d\lambda$ (as well as equation \ref{epsilonderivative}) tends to zero with $\epsilon_\lambda$ approaching a constant ``stretching" term, thus nulling the second term in equation \ref{shape}. From the surviving first term, we can see clear degeneracies between the stellar contamination $\epsilon_\lambda$ and the scale height $H$ which both act like ``stretching" terms on the opacity structure; e.g., an unaccounted stretch due to the spot could be compensated with scale height, leading to a bias in the unknown terminator temperature and/or molecular weight. 

At shorter wavelengths on the other hand, both terms in equation \ref{shape} survive, resulting in a complex shape interaction. The scaling behavior from the first term persists and a power-law like behavior from the second term, yielding additional degeneracies with $z_\lambda$--the wavelength-dependent radius alongside the factors that contribute to the shape of the stellar contamination spectrum, $\frac{dQ_\lambda}{d\lambda}$ i.e the temperature contrast between T$_p$ and T$_s$ and fractional coverage in area of spots, f$_s$.
 
Figure \ref{contamplot} shows example contamination spectra for both spots and faculae with various temperature contrasts and covering fractions (e.g., similar to \citealt{rackham2017,zhang2018}) for realistic stellar spectra (using PHOENIX models from \citealt{allard2003,allard2007,allard2009}). 
 Consistent with the preceeding analytical result, with realistic spectra we also see that increasing the spot covering fraction f$_s$ from a small 0.9$\%$ to a large 12$\%$ value, produces a scaling effect at longer wavelengths. Moreover, at shorter wavelengths, we see a power-law (or slope) effect, similar to what may be expected from scattering in the sub-Neptune atmosphere, as also demonstrated for the case of hot Jupiter HD 189733b \citep{mccullough2014}.
 
 

Now armed with a first order intuition for the effect of stellar contamination on the transmission spectrum--we can explore, numerically through retrievals, the influence of spot contamination.

\section{EXPLORING THE IMPACT OF UNOCCULTED STAR SPOTS ON RETRIEVED ATMOSPHERIC PROPERTIES}\label{sec:retrieval}
In this section we aim to explore the role of spot contamination by addressing the following questions:
\begin{itemize}

    \item Will there be a bias in JWST results if we do not account for stellar contamination in atmospheric retrieval models? How can we rigorously explore this bias to decode degeneracies that arise in transmission spectra?

    \item Does {\it a priori} knowledge of stellar contamination parameters reduce bias in our interpretations? 
    
\end{itemize}

We first describe our retrieval setup, followed by a numerical exploration addressing each of the aforementioned questions.

\subsection{Model Description}
We use the planetary transmission/retrieval tool {\tt CHIMERA}\footnote{\url{https://github.com/ExoCTK/chimera}} \citep{line2013,swain2014,kreidberg2014,kreidberg2015,mai2019} to simulate and retrieve upon JWST-like spectra from 0.6 to 5 $\mu$m (e.g., NIRISS, NIRCam, or NIRSpec). The specific model used here is from \citealt{mai2019} but with modifications to accommodate for the transit light source effect \citep{apai2018,rackham2018} arising from unocculted star-spots (e.g., equation \ref{contam_fac}).  The relevant model variables are given in Table \ref{table_init}. Briefly, the model\footnote{Dropbox link: \url{https://www.dropbox.com/sh/etnftxadm4ho5hj/AAB2EonMGhFLJ97aquHbaeMRa?dl=0}} computes a limb transmission spectrum via the relations in \citealt{brown2001} and \citealt{tinetti2012water}.  We include correlated-K opacities (derived from the line-by-line cross-sections of \citealt{Freedman2008,Freedman2014}) for H$_2$-H$_2$/He collision-induced absorption, H$_2$O, CH$_4$, CO, CO$_2$, NH$_3$, Na, K, TiO, VO, C$_2$H$_2$, HCN, H$_2$S and FeH.  The retrieval uses the ``chemically-consistent" approach (\citealt{kreidberg2015,Kreidberg2018}) given a metallicity and C/O assuming equilibrium gas+condensate chemistry (computed using the NASA Chemical Equilibrium with Applications, CEA, \citealt{gordon1996nasa})), computed along the 3-parameter analytic T-P profile (\citealt{guillot2010, lin13a}). Hazes and clouds are parameterized with a simple ``power-law" \citep{lecav2008} and gray uniform opacity, respectively.




We use a grid of PHOENIX \citep{ allard2003,allard2007,allard2009}
stellar models (from the STScI {\tt pysynphot} routine, \citealt{pysynphot}) in effective temperature to construct the photospheric as well as spot/facular region spectrum (fixing the stellar metallicity to solar and log(g) to 5.0). Within the forward model/retrieval, the photosphere/spot spectra are interpolated to the appropriate effective temperature using Scipy package {\tt griddata} \citep{scipy2019}.

With this modeling framework, we investigate the influence of stellar contamination on a generic M-dwarf orbiting sub-Neptune (a clear 40$\times$ solar, and a cloudy 300$\times$ solar atmosphere, Table \ref{table_init}) observed with a JWST-like platform from 0.6-5 $\mu$m (broken up between NIRISS-like: 0.6-2.5 $\mu$m and NIRCam/NIRSpec G395-Like: 2.4-5.0 $\mu$m.).  In all cases the resolving power (R) is assumed at 100 and for simplicity, we assumed a wavelength independent uncertainty of 30 ppm per bin--loosely motivated by reasonable JWST performance expectations (e.g., \citealt{greene16}). We do not randomize the data-points so as to remove any bias due to arbitrary noise instantiations (e.g.,  \citealt{feng2018} and \citealt{mai2019}).  Parameter estimation and model selection (used to assess the need for spot-correction) are performed with the {\tt PyMultiNest} \citep{Buchner2014,Feroz2009} nested sampling \citep{skilling2006} routine with uniform priors, given in Table \ref{table_init}. 


Figure \ref{scenarioplot} illustrates the influence of unocculted spot-contamination on sub-Neptune transmission spectra. Consistent with recent studies \citep{rackham2018,pinhas2018}, there can be a significant shape change in the transmission spectrum, where the wavelength-dependent stellar contamination parameters could mimic atmospheric effects on the planet's spectrum (e.g., molecular features and haze-like power-law slopes).

\subsection{To what degree is there bias if not accounting for unocculted star spots?}\label{biassection}
\subsubsection{JWST NIRISS-like Bandpass}

\begin{figure*}[!tbp]
    \includegraphics[width=\textwidth]{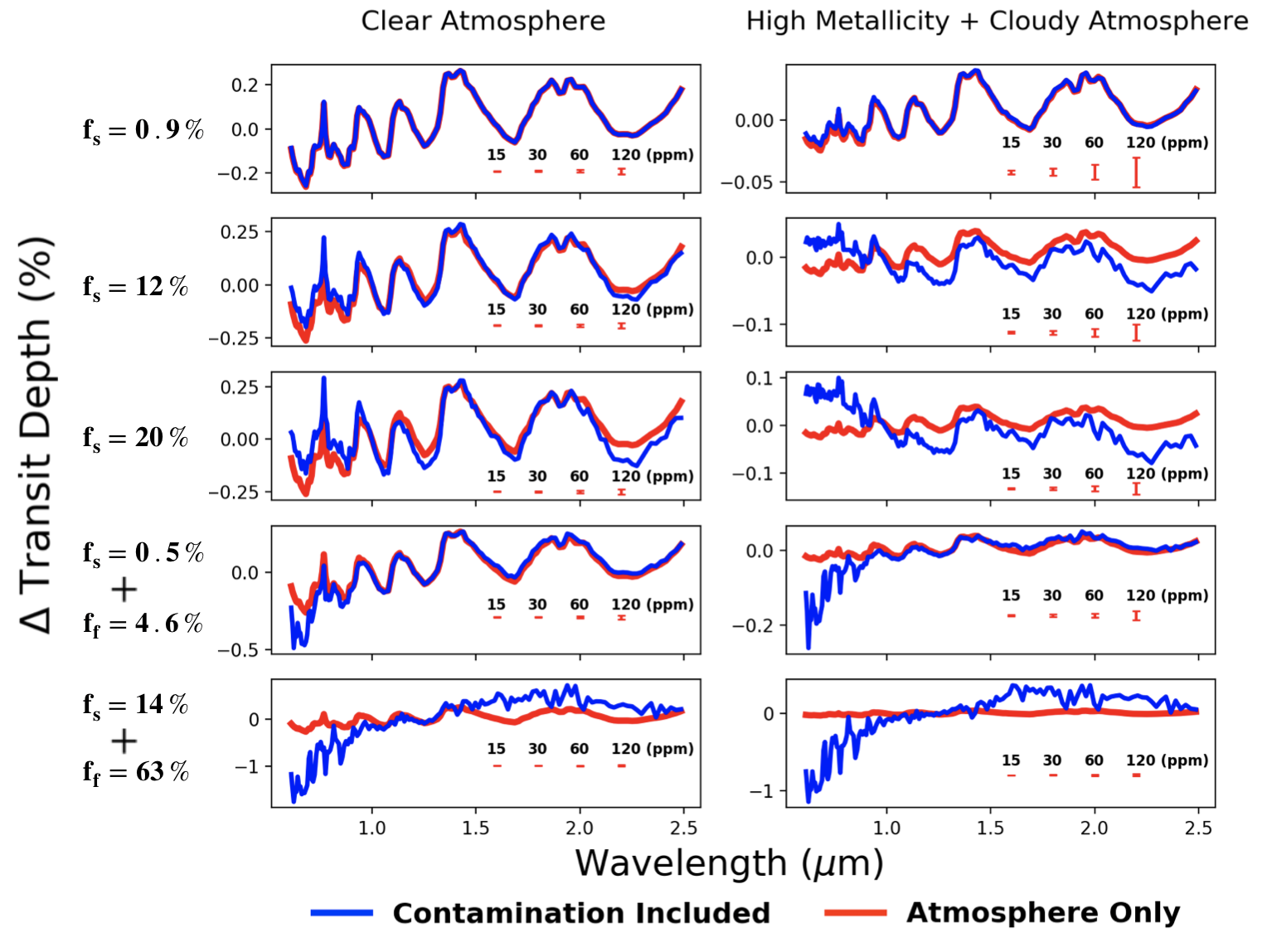}
    \caption{Transmission spectra resulting from including (blue) and ignoring (red) corrections for stellar photosphere with varying degrees of heterogeneities on the host M-dwarf (rows) under a solar composition atmosphere (left) and a cloudy, high metallicity (300$\times$ Solar) atmosphere (right). Representative spectral error bars (15-120 ppm) are shown in the corner of each panel. We see here that contamination covering fractions less than 1$\%$ do not produce a significant shape change in the transmission spectrum.} 
    \label{scenarioplot}
  \end{figure*}

We choose spot/faculae covering fractions and contrasts to be consistent with 1$\%$ I-band rotational variability for mid-to-late M-dwarfs \citep{newton2016,rackham2018}. Specifically, We have assumed T$_s$ = 0.86 x T$_p$ K (or C$_s$ = 0.86) and T$_f$ = T$_p$ + 100 K (or C$_f$ = (T$_p$+100)/T$_p$) \citep{gondoin2008,afram2015,rackham2018}. 
We also explore a range of planetary atmospheric conditions (e.g., clear, cloudy, high metallicity (300$\times$ solar)/molecular weight) given in Table \ref{table_scenarios}  and shown in figure \ref{scenarioplot}.

 \begin{figure*}[!tbp]
    \includegraphics[width=\textwidth]{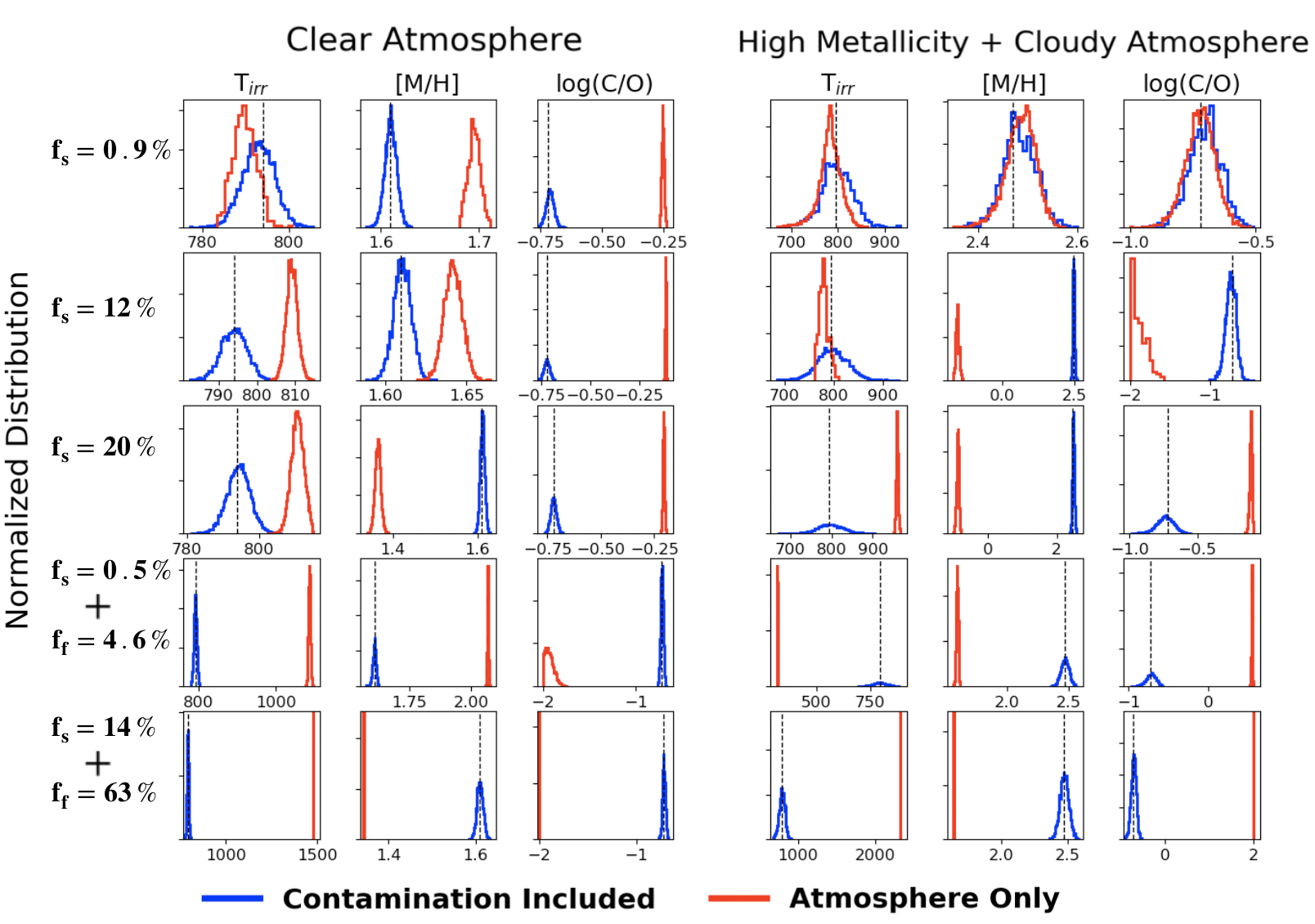}
    \caption{Constraints and biases on T$_{irr}$, [M/H], and log(C/O) retrieved from the transmission spectra (0.6-2.5$\mu$m, R=100, 30 ppm precision) of a clear (left) and cloudy high metallicity (right) sub-Neptune both accounting for (blue) and not accounting for (red) spot contamination for a spectrum constructed with contamination. Each row is labeled with the contamination properties used to construct the ``true" model.  The vertical dashed line indicates the ``true" parameter values. An increasing degree of bias in the model that does not account for contamination (red) is seen as the spot/faculae covering fraction increases (top-to-bottom). A significant bias is present in all clear-atmosphere scenarios, though covering fractions less than 1$\%$ result in little bias for a cloudy/high metallicity scenario (right).} 
    \label{histoscenarioplot}
  \end{figure*}

\begin{figure*}[!tbp]
    \includegraphics[width=\textwidth]{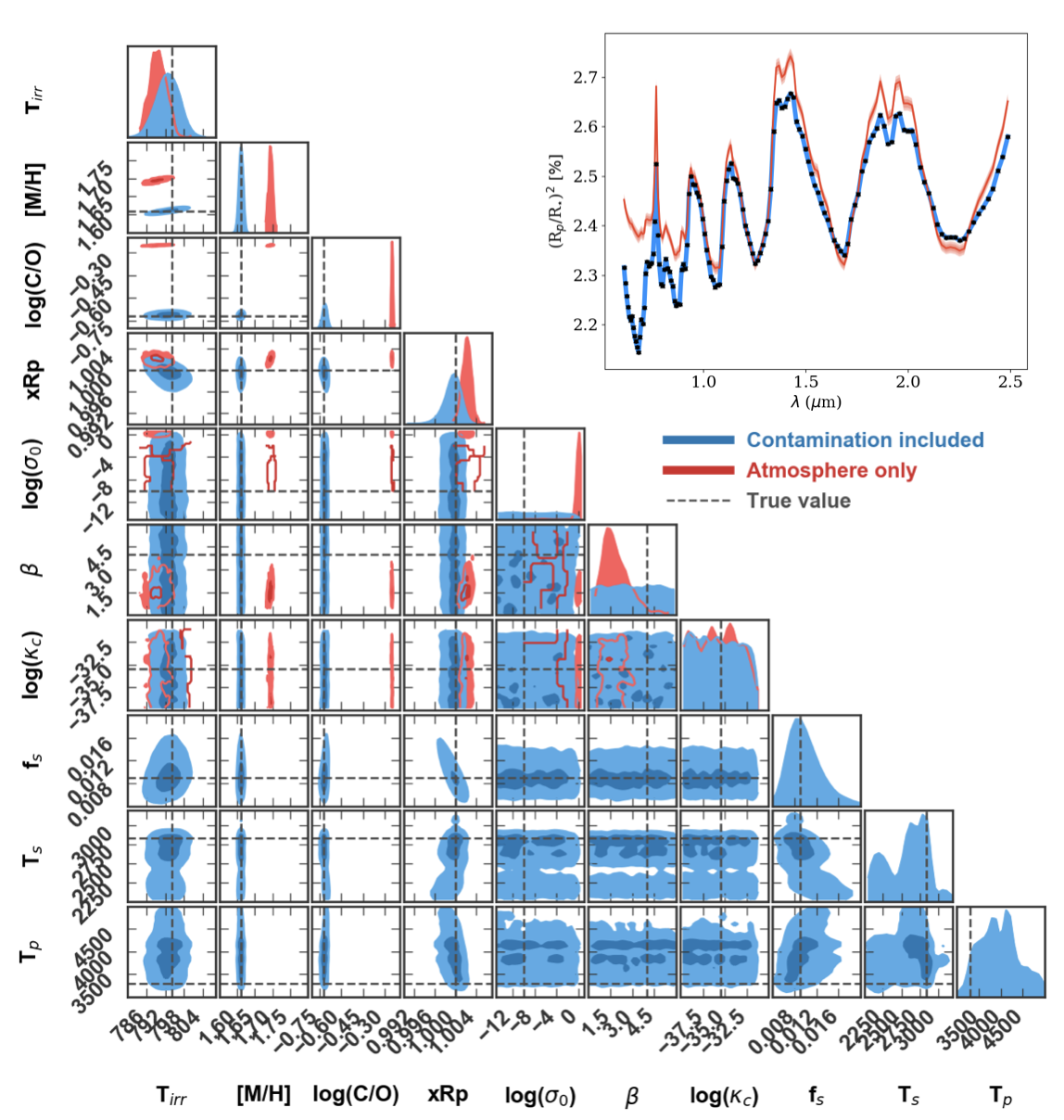}
  \caption{Corner plot summarizing the constraints and degeneracies from a NIRISS-like transmission spectrum (30 ppm sensitivity) for a clear, 40$\times$ solar composition sub-Neptune orbiting an M-dwarf with a 0.9$\%$ coverage (no faculae) when accounting for spot contamination (blue) and not accounting for it (red). There is significant bias in the inferred atmospheric properties (red vs blue) as well as mismatch in the spectral fit when excluding the spot parameterization.}
  \label{clear_pairs_plot}
 \end{figure*}

For each scenario in Table \ref{table_scenarios} (as observed from 0.6-2.5 $\mu$m, R=100, with 30 ppm precision) we  perform a series of retrievals---first a control case that assumes a homogeneous stellar photosphere (no spot contamination correction, retrieves only the atmospheric properties T$_{irr}$, [M/H], log(C/O), log($\sigma_0$), $\beta$, and log($\kappa_c$)) and second accounting for the spot contamination (e.g., additionally retrieves for T$_p$, T$_s$, f$_s$, T$_f$, and f$_f$ ). Two figures of merit are used when assessing the level of ``bias" in the inferred planetary properties. The first (figures \ref{histoscenarioplot} and \ref{clear_pairs_plot}), a qualitative metric describing the bias by simply measuring the actual deviation of the posterior probability distributions (medians of the distributions) from the truth values.  We only report the bias in terms of the width of the posterior probability distributions (i.e. shift of the histogram in units of $\sigma$ width relative to the retrieved precision) for a select few cases with distributions lying fully within the chosen prior range. Instances where the posterior histograms converge partly on the edge of the priors skew the true level of bias by falsely portraying additional width to the histograms---therefore, this is purely a ``qualitative metric'' to measure the bias in the inferences (e.g. see figure \ref{histoscenarioplot}, right column for high metallicity and cloudy planetary atmosphere with f$_s$ = 12$\%$ for the carbon to oxygen ratio--log(C/O)) . 
The second metric (to be subsequently discussed) is a quantitative measure of the level of bias in the inferred planetary parameters, accounted for by computing the Bayes factor (Table \ref{table_scenarios}, last column) between the atmosphere only and the atmosphere+spot model retrieval results for all the cases with varying levels of stellar heterogeneities. 


In all cases (see Table \ref{table_scenarios})\footnote{{As a sanity check, we also performed a similar exercise on an uncontaminated stellar photosphere to determine whether the introduction of the contamination parameterization within the retrieval would unduly influence the retrieved atmospheric properties. Unsurprisingly, we found that it did not, e.g., no ``intrinsic" bias is introduced by adding in the contamination parameterization.}} with a JWST NIRISS-like transmission spectrum (covering 0.6-2.5 $\mu$m, with 30 ppm precision) of a generic sub-Neptune orbiting an M-dwarf, we demonstrate that there is significant ``qualitative bias'' in the retrieved atmospheric properties when not correcting for stellar contamination. For instance, a case with a spot covering fraction f$_s$~=~0.9$\%$, an uncorrected transmission spectrum for a clear atmosphere sub-Neptune yields a bias in the retrieved planetary irradiation temperature (T$_{irr}$), bulk metallicity ([M/H]), and the bulk carbon-to-oxygen ratio (log(C/O)) of 1.6$\sigma$, 13.7$\sigma$, and 95.5$\sigma$ respectively (illustrated in figures \ref{histoscenarioplot}
 and \ref{clear_pairs_plot}). Moreover, in figure \ref{clear_pairs_plot}, a correlation between the planetary radius (fiducial radius at 10 bar--$\times$R$_p$) and the stellar spot covering fraction (f$_s$) emerges from the retrieval 2-D posterior histograms. Consistent with the previously illustrated analytical result (Equation \ref{shape}), the wavelength dependent stellar contamination term is degenerate with both the planetary atmospheric scale height $H$ as well as the wavelength dependent sharp occulting disk radius $z_\lambda$ altogether affecting the shape of the transmission spectrum. We also see this effect in figure \ref{scenarioplot}, where increasing the level of stellar photospheric heterogeneity (a linear combination of the spot covering fraction and the temperature contrast between the spot and the immaculate photosphere) modifies the shape of the transmission spectrum and hence the derived planetary radius significantly, beyond anticipated spectral precisions (15 ppm - 120 ppm). Focusing, however, only on 30 ppm precision as an example case, the trend of increasing bias with the first metric (deviance from truth) persists for a combination of stellar heterogeneities, ranging from 0.5$\%$ to 20$\%$ in spot covering fraction and 4.6$\%$ to 63$\%$ for faculae covering fraction, as illustrated in figure \ref{histoscenarioplot}.

\begin{figure*}[!tbp]
    \includegraphics[width=\textwidth]{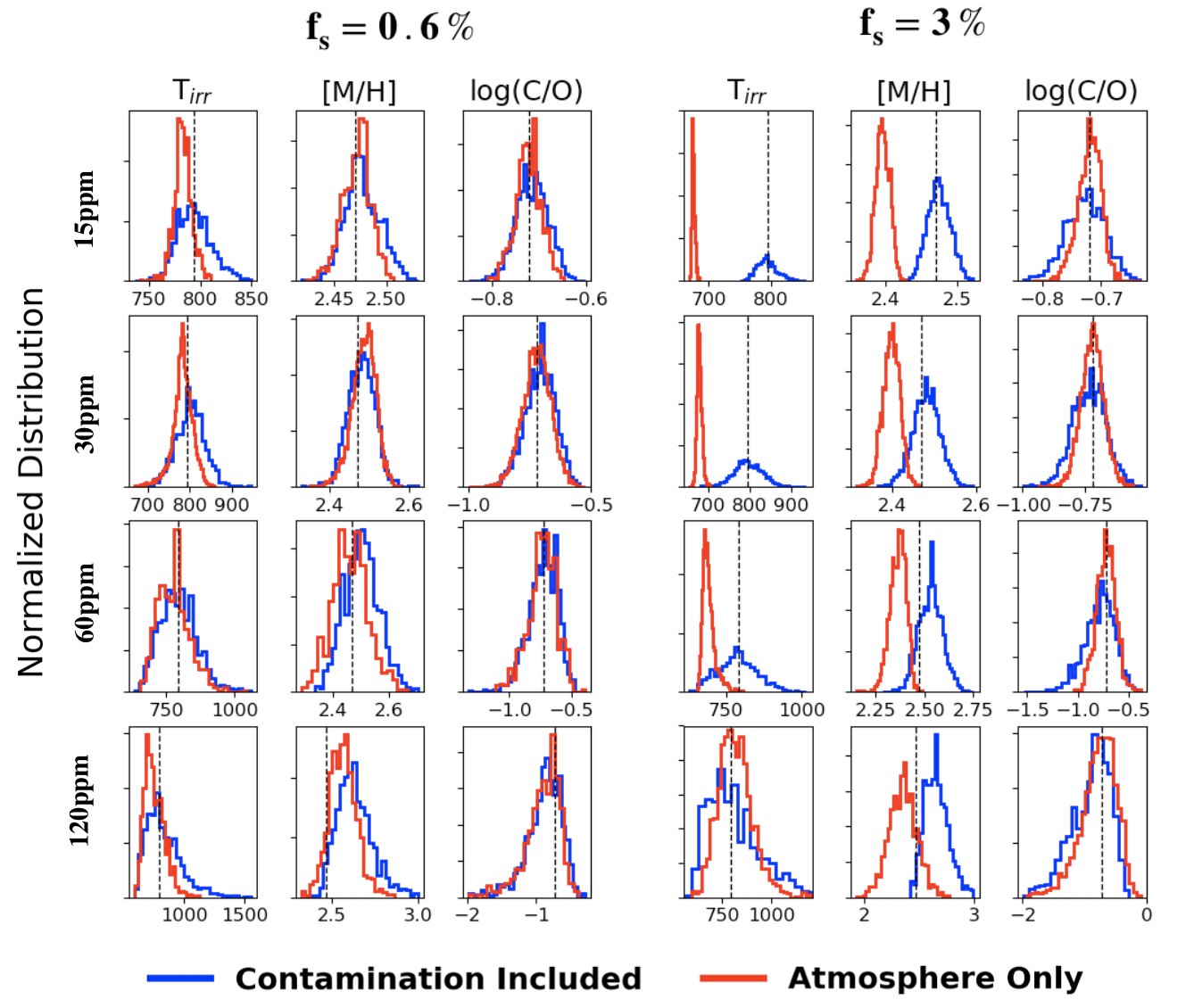}
  \caption{Constraints/bias on T$_{irr}$, [M/H], and log(C/O) as a function of spectral precision (15-120 ppm, rows) and two spot covering fractions (0.6$\%$ left, 3$\%$ right) under the high metallicity cloudy atmosphere setup, as observed with a NIRISS-like instrument. The red curves are from the atmosphere only model (no spot correction) and the blue for the spot retrieved model. Generally, increasing the spot covering fraction increases the level of bias but over all bias decreases with larger spectrophotometric uncertainties on the transmission spectrum.}
  \label{differr_multihisto}
\end{figure*}

We also find that for the smallest heterogeneity fractions, muting the transit spectral features with clouds and high metallicity reduces the effective ``qualitative bias" as seen in the retrieved planetary parameters (see figure \ref{histoscenarioplot}, top right panel).  For f$_s$ = 0.9$\%$ in particular, we find that there is a bias of 0.48, 0.55 and 0.01 $\sigma$ for the inferred irradiation temperature, planetary metallicity, and carbon-to-oxygen ratio respectively, for an uncorrected transmission spectrum. With increasing stellar heterogeneity covering fractions however (ranging from f$_s$ = 12$\%$ to 20$\%$  alone and for f$_s$ = 0.5$\%$ along with f$_f$ = 4.6$\%$ up to 63$\%$), the qualitative increase in the bias (deviation from the truth values) of the posterior probability distributions for cloudy and high-metallicity atmospheres (although slightly reduced) are consistent with those of a clear sub-Neptune atmosphere. Only for small spot covering fractions, we find that the level of bias falls well within 1$\sigma$ of the true value, and this effect is persistent through a range of spectral precisions (15 - 120 ppm, see figure \ref{differr_multihisto}, left column for f$_s$ = 0.6$\%$). This is primarily because; as the transit feature signal-to-noise decreases (whether due to clouds and/or high mean molecular weight) thereby increasing degeneracy in the transmission spectrum--results in $larger$ parameter uncertainties. These uncertainties therefore, seemingly reduce the effective bias under our first figure of merit, in turn motivating the question: {\it at what degree of stellar heterogeneity in fractional coverage does the bias in retrievals matter?}  

To address this question, we start by qualifying our second figure of merit--the Bayes factor; which provides a quantitative measure of the need to include as ``nuisance" parameters, a stellar correction scheme (last column in Table \ref{table_scenarios}, for 30 ppm precision). Specifically, we compute the log of the Bayes factor ($ln B$, \citealt{cornish2007}) which is a metric permitting quantitative comparison of two models with different complexity (e.g., one with and without a spot-correction parameterization). For all cases in Table \ref{table_scenarios}, we highlight two points illustrating the behavior of the Bayes factor: (1) $ln B$ increases with an increase in contamination covering fraction and (2) decreases with increasing cloudiness/molecular weight in the planetary transmission spectrum. The first point is consistent for all planet atmosphere scenarios, as seen previously with our first figure of merit (figure \ref{histoscenarioplot}). For instance, a clear atmosphere scenario with increasing spot covering fraction f$_s$ from 0.9 to 20$\%$ increases ln$B$ from 10 to 10,634, with the entire range demonstrating strong evidence for preferring the model including stellar contamination as per the Jeffrey's scale \citep{trotta2008}. The increasing trend in ln$B$ is similar through all atmospheric scenarios including for cases with cloudy and high-metallicity atmospheres, where the ln$B$ ranges from 0.23 to 2479 for f$_s$ increasing from 0.9 to 20$\%$ (see Table \ref{table_scenarios}). This pattern is consistent for combinations of spot and faculae covering fractions both for clear as well as cloudy and high-metallicity atmospheres, thereby quantitatively indicating the degree to which the model including stellar contamination is preferred. The second point shows a decrease with increasing feature muting (due to clouds/metallicity). For the entire range of stellar contamination cases in Table \ref{table_scenarios} we see this behavior, especially for the case of a small spot covering fraction of 0.9$\%$ (under the cloudy and high-metallicity atmosphere), which yields an ln$B$ = 0.23 demonstrating inconclusive evidence favoring either model (true for all ln$B$ values on the Jeffrey's scale $<$ 1, \citealt{trotta2008}). Despite the decreasing trend of ln$B$ with increasing atmospheric degeneracies, for other spot and spot+faculae covering fraction combinations, the Bayes factors strongly prefer the model including stellar corrections to prevent bias in the inferences, i.e. ln$B$ $>$ 5. From this trend, we can say that regardless of the planetary atmosphere properties explored here, a system with a host star of spot covering fraction $>$ 1$\%$ must include a parameterization for stellar activity to mitigate the bias atmospheric inferences.

\begin{figure}[!tbp]
    \includegraphics[width=\columnwidth]{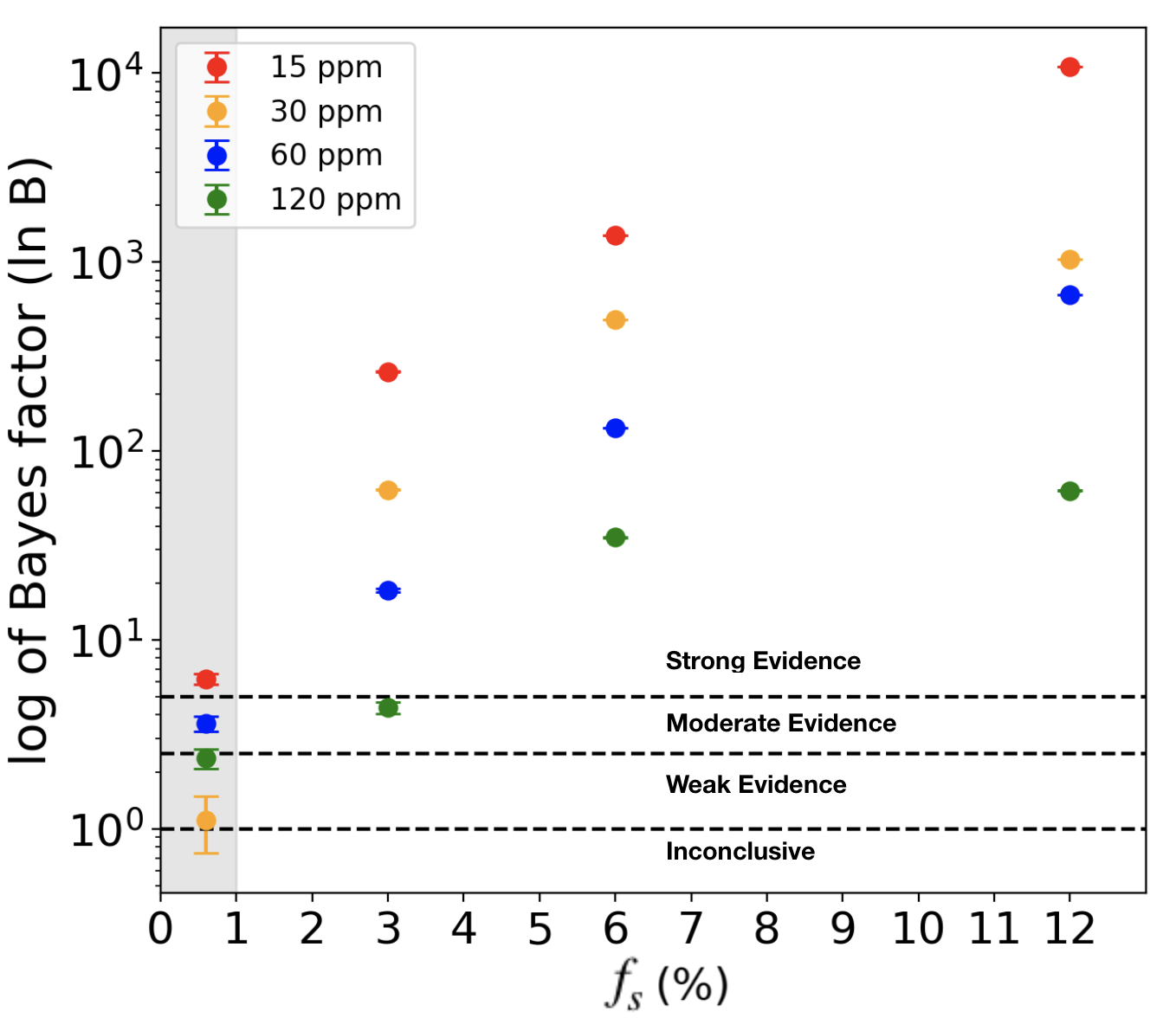}
  \caption{Bayes factor behavior between the atmosphere only and atmosphere+spot correction model for a spot contaminated cloudy high-metallicity atmosphere as a function of spot covering fraction (horizontal-axis) and spectrophotometric precision over a NIRISS-like pass-band (colored dots).  The horizontal black dotted lines indicate the degree to which the contamination included model is preferred over the atmosphere only model \citep{trotta2008}. Above 1$\%$, the corrected model is preferred mostly with strong evidence for all precision levels with the exception of 120 ppm error bar case for  3$\%$ coverage. Below 1$\%$ coverage the bias is not significant except for the 15 ppm precision however; where the stellar correction model is always highly preferred. Note, the awkward crossing of the 30ppm point below 1$\%$ is due to the inaccuarcies in computing Bayes factors in low evidence regimes (e.g., \citealt{Lupu16}).}
  \label{bayesplot}
\end{figure}

We now focus on quantifying the bias as a function of spectral precision, under the more likely scenario of a cloudy-high metallicity atmosphere \citep{fortney2013}. We do this by generating a sparse grid of retrievals--varying spot covering fractions on the first dimension, ranging from 0.6$\%$ to 12$\%$ and spectral precision on the second dimension ranging from 15 to 120 ppm, consistent with expectations for JWST based on achieved HST/WFC3 precisions (e.g., down to 15 ppm, \cite{line2016}).  For each point on the grid, we perform two sets of retrievals---the first with the atmosphere only model and the second with stellar corrections included. Our main findings from this exercise are: for the entire range of precisions, increase in fractional coverage in area of spots (f$_s$) on the stellar photosphere increases the significance with which the model with stellar corrections is favored (ln$B$). Particularly for spot covering fractions f$_s$ $>$ 1$\%$, we find that the Bayes factor ranges from moderately favoring the model with stellar contamination included (e.g. case of 120 ppm precision at f$_s$ = 3$\%$ in figure \ref{bayesplot} is in the moderate evidence regime as per the Jeffrey`s scale) to strongly favoring the model for all other sensitivities. This trend provides strong evidence for the spectral presence of stellar heterogeneities in the transmission spectrum throughout these levels of precisions.


For cases of f$_s$ $<$ 1$\%$ and for all sensitivities $>$ 30 ppm on our grid (see figure \ref{bayesplot}), we only see weak to moderate evidence for stellar contamination. This is consistent with the reduction in our ``qualitative bias'' metric (see figure \ref{histoscenarioplot}, top panel, right column and figure \ref{differr_multihisto} left column) as demonstrated previously, due to increasing the level of degeneracies in the spectrum for cloudy/high metallicity scenarios, especially at small spot covering fractions. For the case of 15 ppm precision however, the uncertainties on the transmission spectrum are significantly small enough that even a region of heterogeneity as small as f$_s$ = 0.6$\%$ shows a very high $ln B$ providing strong spectral evidence for the contamination. Although, this is not apparent from our ``qualitative" bias metric as seen in the top left panel of figure \ref{differr_multihisto} for f$_s$ = 0.6$\%$.
Overall, we see that spot covering fractions $>$ 1$\%$, for our defined grid of precision levels, increasing f$_s$ leads to an increase in the ``quantitative bias" ($ln B$) providing strong evidence for stellar contamination, however this increase is penalized by the level of uncertainties on the JWST transmission spectrum, thereby reducing the apparent ``qualitative bias" (deviance from the truth) as we see in the right column of figure \ref{differr_multihisto}.

\begin{table*}[t]
\renewcommand{\thetable}{\arabic{table}}
\centering
\caption{Model paraemters, nominal ``truth" values, and uniform prior ranges.}\label{table_init}
\begin{tabular}{| l | l | l | l | l |}
\hline
Parameter & Description & Initial Value & Status & Prior Range\\
\hline
\decimals
\hline
log($\kappa_{ir}$)$^a$ & Thermal profile gray IR opacity (cm$^2$/g) & 0.306 & Fixed & N/A\\
log($\gamma_v$)$^a$ & Thermal profile Vis/IR opacity & -2.02 & Fixed & N/A\\
R$_p^b$ & Adopted 10 bar Planetary Radius (R$_{jup}$) & 0.239 & Fixed & N/A\\
M$_p^b$ & Planetary Mass (M$_{jup}$) & 0.0204 & Fixed & N/A\\
R$_*^b$ & Stellar Radius (R$_\odot$) & 0.211 & Fixed & N/A\\
\hline
\hline
T$_{irr}^a$ & Irradiation Temperature (K) & 794. & Retrieved & [300,3000]\\

[M/H] & Planetary Metallicity & 1.61 (40$\times$) or 2.47 (300$\times$) & Retrieved & [-2,3]\\
log(C/O) & C-to-O ratio & -0.72 & Retrieved & [-2,2]\\
$\times$Rp & 10 bar Radius Scaling & 1. & Retrieved & [0.5,1.5]\\
log($\sigma_0^c$) &  Haze cross section amplitude  & -10. & Retrieved & [-15,2]\\
             
$\beta^c$ & Haze Scattering Slope & 4. & Retrieved & [0,6]\\
log($\kappa_c$) & Well-mixed gray cloud opacity & -35.5 (clear) or -29 (cloudy) & Retrieved & [-40,-20]\\
T$_{p}$ & Stellar Photosphere Temperature (K) & 3300 & Retrieved & [2000,4000]\\
T$_{s}$ & Stellar Spot Temperature (K) & 2838 & Retrieved & [2000,4000]\\
f$_s$ & Fractional Coverage of Spots & Case Specific$^d$ & Retrieved & [0,1]\\
T$_f$ & Stellar Faculae Temperature (K) & 3400 & Retrieved & [2000,4000]\\
f$_f$ & Fractional Coverage of Faculae & Case Specific$^d$ & Retrieved & [0,1]\\
\hline
 \multicolumn{4}{l}{%
  \begin{minipage}{15.5cm}
$^a$ 3-Parameter model for thermal profile \citep{guillot2010}\\
$^b$ GJ1214/b, from exoplanets.org \citep{exoplanet.org}\\
$^c$ From \citealt{lecav2008}.  Haze cross section given by $\sigma_{\lambda}=\sigma_0 (\lambda_0/\lambda)^{\beta}$ w ith $\lambda_0$=0.43 $\mu$m and $\sigma_0$ given in units relative to H$_2$ rayleigh scattering (2$\times 10^{-27}$ cm$^{2}$)\\
$^d$ Initial Values are specific to cases of spot and faculae covering fractions, listed in Table \ref{table_scenarios}, also given in \citealt{rackham2018,zhang2018}
  \end{minipage}%
}\\
\hline
\end{tabular}
\end{table*}

\begin{table*}
\renewcommand{\thetable}{\arabic{table}}
\centering
\caption{Bayes Factor between an atmosphere only model and a model that accounts for the spot contamination under the several atmosphere scenarios (Table \ref{table_init}) and a NIRISS-like (30 ppm, R=100, 0.6 - 2.5$\mu$m) and NIRCam-like (2.4-5 $\mu$m) observational setup.}\label{table_scenarios}
\begin{tabular}{|l | l | l |}
\hline
Stellar Contamination  & Atmosphere Type & Bayes factor (ln B)\\
\hline
NIRISS-like  &  &  \\
\hline
\decimals
\hline
f$_s$ = 0.9 $\%$ & Clear  & 10   \\
   & Cloudy &  2.5 \\
   & High Metallicity + Cloudy &  0.23   \\
\hline
f$_s$ = 12 $\%$ & Clear & 3191   \\
   & Cloudy &  2653  \\
   & High Metallicity + Cloudy & 1036 \\
\hline
f$_s$ = 20 $\%$ & Clear & 10,634   \\
   & Cloudy &  6,085   \\
   & High Metallicity + Cloudy & 2,479\\
\hline
f$_s$ = 0.5 $\%$, f$_f$ = 4.6 $\%$ & Clear &  14,634   \\
   & Cloudy & 14,628\\
   & High Metallicity + Cloudy & 6,389   \\
\hline
f$_s$ = 14 $\%$, f$_f$ = 63 $\%$ & Clear & 1,575,792 \\
   & Cloudy & 1,543,800\\
   & High Metallicity + Cloudy &  66,478   \\
\hline
NIRCam-like  &  & \\
\hline
f$_s$ = 14 $\%$, f$_f$ = 63 $\%$   & Cloudy &  63,295\\
\hline
\hline
\end{tabular}
\end{table*}

\subsubsection{JWST NIRCam/NIRSPEC-G395-Like Bandpass}
In section \ref{sec:analytical} we showed that stellar contamination has a stretching effect on the shape of the transmission spectrum at longer wavelengths, and was therefore degenerate with the scale height. In this section we explore the role of stellar contamination over the JWST NIRCam/NIRSPEC (e.g., G395)-like band pass (2.4-5.0 $\mu$m, and again assuming a constant 30 ppm precision) for a cloudy sub-Neptune atmosphere. The 3-5 $\mu$m region is critical for determining the carbon inventory in planetary atmospheres as large CO, CO$_2$ and CH$_4$ features are present throughout.  We explore a scenario with a 14\% spot and 63\% faculae fraction, under the same temperature contrast as above.

\begin{figure}[!tbp]
    \includegraphics[width=\columnwidth]{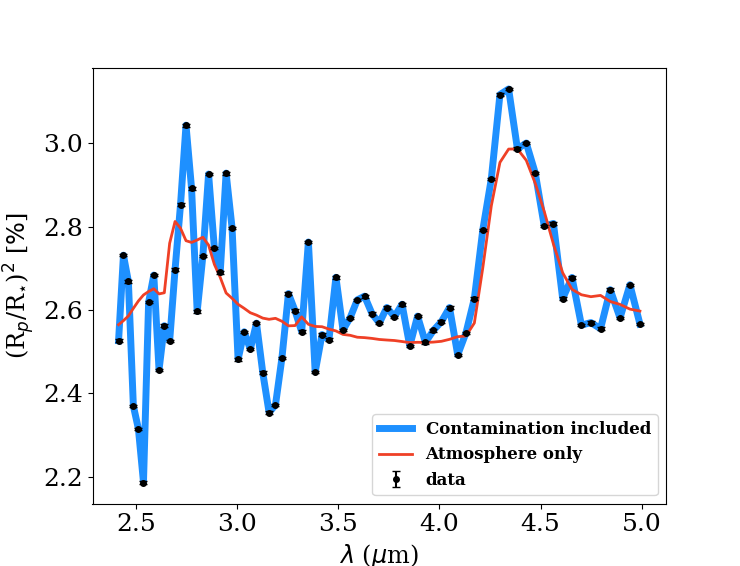}
  \caption{Spot contamination influence over a NIRCam/NIRSpec G395-like band pass with 30 ppm precision for a cloudy sub-Neptune atmosphere with spot and faculae covering fractions of 14$\%$ and 63$\%$, respectively. The true spectrum is shown as the black points (errors are small), the best fit model that includes the contamination parameterization is shown in blue, and the model that fails to account for contamination (atmosphere only) in red.  Most of the ``high-frequency" spectral features seen in this wavelength range under this particular spot/faculae setup are due to the stellar contamination.  The strong 4.25$\mu$m CO$_2$ feature at high metallicity stands above the stellar contamination ``noise".}
  \label{nircamspec}
\end{figure}
\begin{figure*}[!tbp]
    \includegraphics[width=\textwidth]{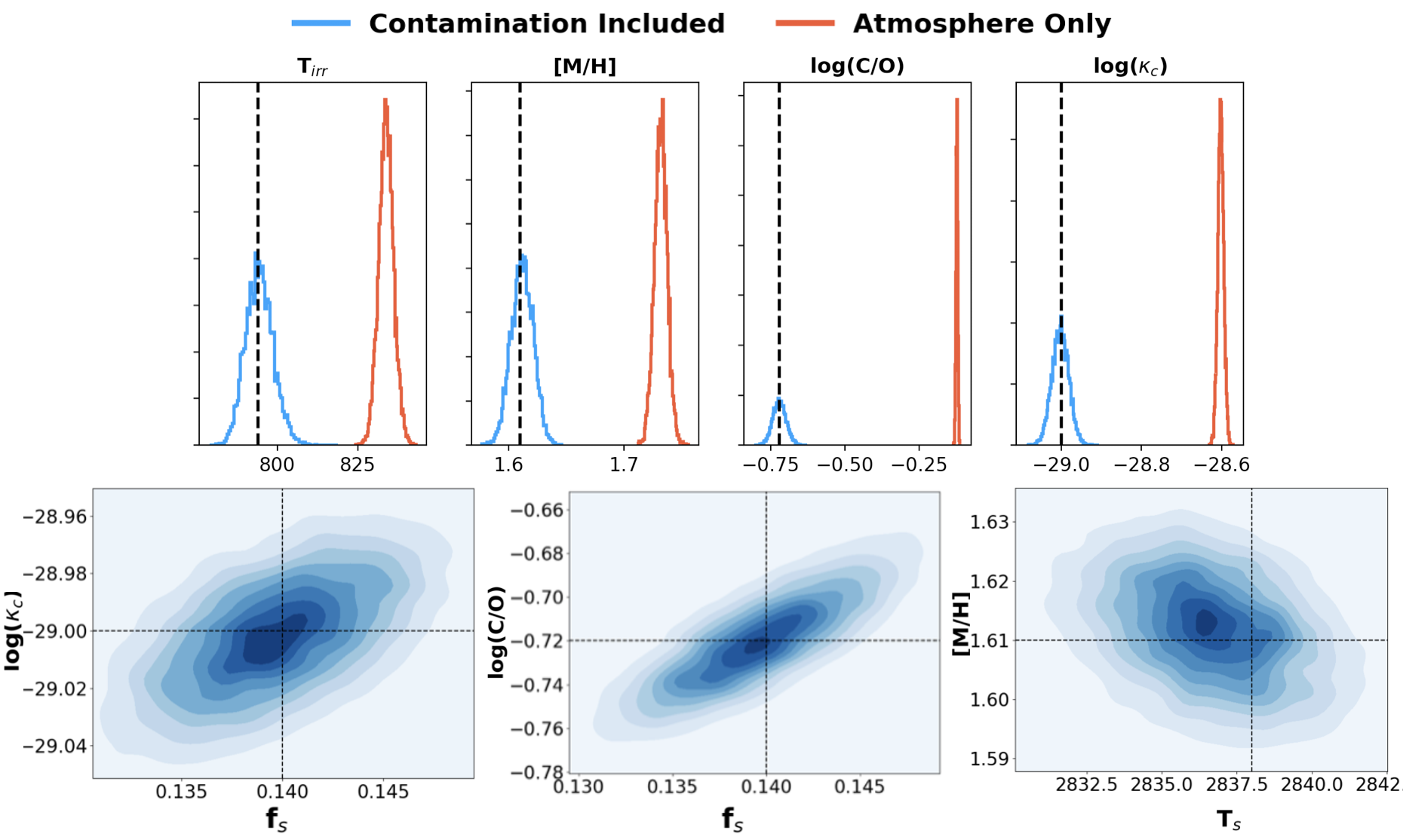}
  \caption{Retrieval Results/biases for the irradiation temperature (T$_{irr}$), metallicity ([M/H]), carbon-to-oxygen ratio (log(C/O)), and cloud opacity (log($\kappa_c$)) over the NIRCam/NIRSpec G395-like wavelength range (and 30 ppm precision) for a cloudy sub-Neptune orbiting a 14$\%$ spot, 63$\%$ faculae contaminated M-dwarf under the atmosphere only scenario. The top row illustrates the constraints and biases on the key atmospheric parameters under the atmosphere only (red) and the contamination correction included scenario (blue).  The bottom shows the 2-D histograms illustrating the prominent degeneracies between the spot parameters f$_s$ and T$_s$ with the atmospheric parameters.}
  \label{nircamplot}
 \end{figure*}

We find that when not including the contamination model parameters there is a deviation of 21.7, 197 and 16.9 $\sigma$ respectively, from the truth values, for the metallicity, C/O ratio, and irradiation temperature along with a log Bayes factor value of 63,295, which falls under the regime of strong evidence (\citealt{trotta2008}) for stellar contamination (Table \ref{table_scenarios}, figures \ref{nircamspec} and \ref{nircamplot}). 
From the 2-D posterior probability distributions (figure \ref{nircamplot}), we also find noteable correlations that emerge between the spot covering fraction (f$_s$) and the  gray cloud opacity ($log(\kappa_c)$) as well as a correlation between log(C/O) and the spot covering fraction. These degeneracies are consistent with our qualitative findings in  section \ref{sec:analytical} (equation \ref{shape}), between cloud opacity and metallicity. We also note in figure \ref{nircamspec} that the stellar contamination spectrum is a major contributor to the variations in the shape of the transmission spectrum within this bandpass.
These results suggest that stellar contamination can have a similarly large influence on longer wavelengths due to the presence of stellar molecular spectral features (see our figure \ref{nircamplot}, also \citealt{rackham2017}).

\subsection{How precisely do we need to know the stellar/spot properties to remove any bias?} 
The natural next issue to arise in light of the influence of spot contamination on retrieved atmospheric properties, is determining the level of required \textit{a-priori} spot contamination knowledge to mitigate the bias. In this section, we investigate whether having any prior knowledge of the level of stellar contamination on the host star's photosphere would mitigate atmospheric retrieval bias.

First, we again simulate a cloudy, high-metallicity sub-Neptune atmosphere with a 12$\%$ region of unocculted spot as observed under the JWST-NIRISS-like bandpass with 30 ppm precision. We choose this spot covering fraction as it falls under the regime that is sensitive to detect stellar contamination bias (f$_s$ $>$ 1$\%$) as demonstrated in section \ref{biassection} and figure \ref{bayesplot}.
For this exercise, we perform retrievals using two modeling approaches: (1) \textit{Joint Fit Model}  which simultaneously fits the stellar spectrum (assuming linear combination of the the photosphere and spot spectra) and transmission spectrum for the contamination parameters under the assumption that the out-of-transit stellar photosphere has similar properties as the in-transit photosphere (e.g. \citealt{zhang2018}), and  (2)\textit{Corrected Transmission Spectrum Model} which only fits for the contamination parameters from the transmission spectrum itself (as in the above sections).
We stress that we implicitly operate under the assumption that the stellar model is an accurate representation of an actual stellar spectrum, e.g., we assume no systematic ``bias" in stellar model fits (however, we explore this assumption in section \ref{checkstellarmod} and figure \ref{starcompareplot}).  We also assume that the absolute flux calibration uncertainty of a stellar spectrum is incorporated into our assumed photometric precision, which may not necessarily be true.

\begin{figure*}[!tbp]
  \includegraphics[width=\textwidth]{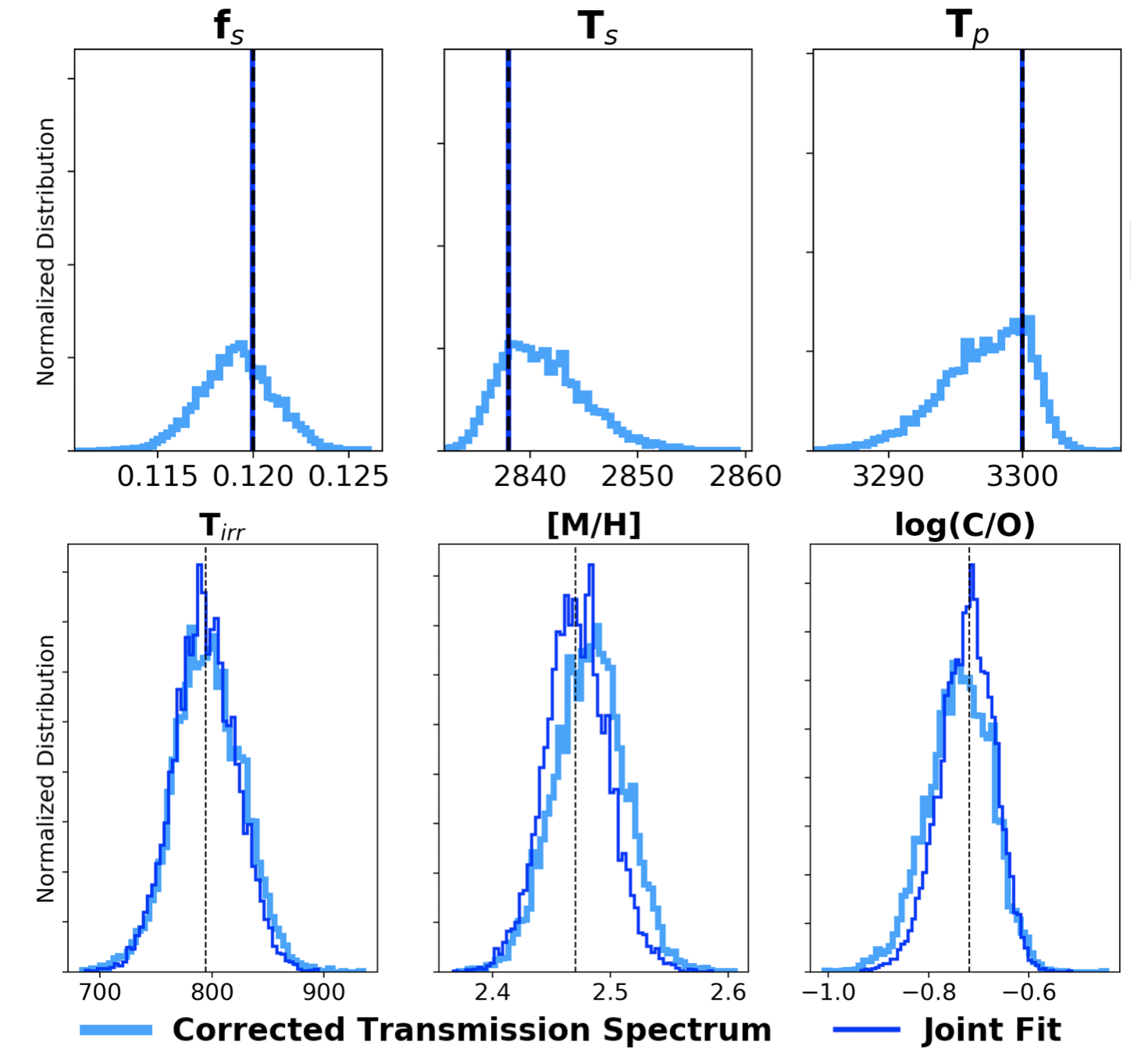}
    \caption{Illustration of the impact of simultaneously retrieving on the stellar spectrum (\textit{Joint Fit Model}) to determine more precisely the spot properties, under the cloudy high metallicity scenario with a 12$\%$ spot covering fraction (3300 K photosphere and 2838 K spot)  as observed under the JWST-NIRISS-like bandpass with 30 ppm precision. The top panel compares the constraints on the stellar contamination parameters when utilizing only the transmission spectrum (light blue) and including both the stellar spectrum and transmission spectrum in the retrieval (dark blue). The bottom panel illustrates the behavior of the atmospheric parameter constraints when including the stellar spectrum in the retrieval. Despite the very precise inferences of stellar contamination parameters from the \textit{Joint Fit Model}, there is no significant improvement in atmospheric parameter constraints beyond simply including and marginalizing over the spot parameters on the transmission spectrum alone (\textit{Corrected Transmission Spectrum Model}).}
    \label{preciseplot}
 \end{figure*}

The posterior probability distributions of $f_s$ (fractional coverage of the spot region on the photosphere), $T_s$ (Temperature of the spot region), and $T_p$ (Temperature of the photosphere) retrieved from the disk-integrated spectrum under both approaches are presented in figure \ref{preciseplot} top panel. We see here that the precision in the inferred stellar contamination parameters improves by several orders of magnitude for f$_s$, T$_s$, and T$_p$ respectively. 

However, utilizing the stellar spectrum itself (the joint fit model) shows no statistically significant improvement in the {\it atmospheric} parameter precisions (i.e., we cannot reject the null hypothesis that both samples are drawn from the same population by performing a 2-sample Kolmogorov-Smirnov test \citep{smirnov1939}, see figure \ref{preciseplot} bottom panel). This important result suggests, at least in this example, that simply including a parameterization for the contamination spectrum, and marginalizing over said parameters within a retrieval is enough to ``remedy" the stellar in-homogeneity problem; no need for \textit{a-priori} stellar contamination knowledge (e.g., as in \cite{pinhas2018}). This result however, may only be true if the stellar models are accurate representations of the true photosphere, which we explore in the next section.


\subsection{What if the stellar models are inaccurate?}\label{checkstellarmod}
In the above analyses we assumed that the stellar models (for both the photosphere and spots) were accurate representations of a true contaminated stellar spectrum.  However, model stellar spectra are not necessarily perfect in all situations \citep{veyette2017}. Here, we explore the impact of an inaccurate stellar model on our ability to remedy spot contamination. We repeat the exercise of retrieving for planetary atmospheric properties of irradiation temperature, metallicity and C-to-O ratio from a contamination corrected transmission spectrum with the same modeling approach as described in section \ref{sec:retrieval}, however we use a different stellar model to incorporate stellar contamination in our simulated data.

\begin{figure*}[!tbp]
    \includegraphics[width=\textwidth]{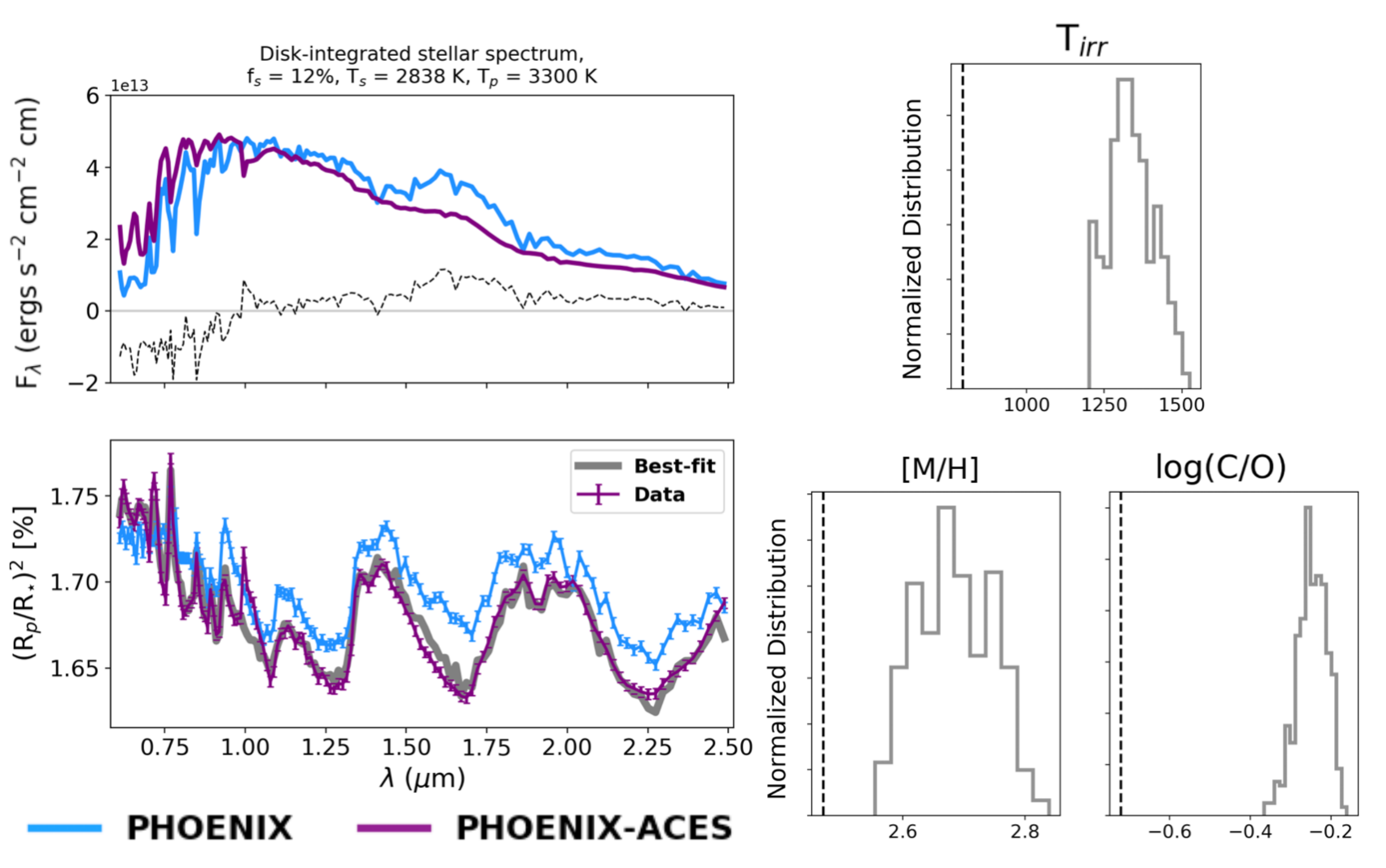}
    \caption{Demonstration of the impact of the choice of stellar model atmosphere used to correct the transmission spectrum, effectively mimicing anticipated deviations of a stellar model with the true stellar spectrum. This is under the cloudy high-metallicity scenario orbiting an M-dwarf with 12$\%$ spot as observed with a 30 ppm precision over NIRISS-like pass-band. We emulate the ``true'' stellar spectrum (top left) and contaminated transmission spectrum (bottom left) with the newer PHOENIX-ACES models \citep{husser2013} (in purple) but apply the correction in the transmission retrieval using the older PHOENIX models \citep{allard2003,allard2007,allard2009} (light blue). In the transmission spectrum panel (bottom left) we compare the ``true spectrum" (e.g., ``data") produced with the PHOENIX-ACES stellar model contamination to the incorrect contamination model produced with the PHOENIX stellar model, along with the ``best" fit PHOENIX corrected stellar model (gray).  The histograms on the right show the strong biases on the atmospheric properties that can occur when an inadequate stellar photosphere model is used for the spot correction (truth values as vertical dashed lines)}
    \label{starcompareplot}
  \end{figure*}

We again simulate a JWST-NIRISS-like transmission spectrum for a cloudy and high-metallicity sub-Neptune  and ``mimic" potential stellar model deviations relative to a true stellar spectrum by using a different stellar model. For the ``true" stellar spectrum, we use the PHOENIX-ACES model (\citealt{husser2013}, with T$_p$ = 3300 K, T$_s$ = 2838 K, f$_s$ = 12$\%$, log(g) = 5.0 and solar metallicity) and for the retrievals, we fit with the older PHOENIX models (used above, also see figure \ref{contamplot}, \citealt{allard2003,allard2007,allard2009}). We are assuming that generational model differences are representative of model--stellar data differences (see Figure \ref{starcompareplot}, top left panel). For this exercise, we assume a spot covering fraction of 12$\%$ as this falls within the regime sensitive to detect bias in the planetary parameter retrievals (f$_s$ $>$ 1$\%$) as we illustrated in Figure \ref{bayesplot}.

The disk-integrated spectra from both our ``true'' star as well as the model star are presented in the top left panel of figure \ref{starcompareplot}. We also see a significant difference in the shape of the same planetary transmission spectrum as expected from the differences translating from their respective contamination spectra (see figure \ref{starcompareplot}). In grey, we show the ``Best-fit'' to our simulated JWST-NIRISS transmission spectrum that includes correction for heterogeneity using the ``true'' stellar spectrum. Overall, we see that the fit is quite reasonable in
explaining the shape of the transmission spectrum, however since the contamination retrieval model is missing the ``true'' stellar information (due to model vs measured stellar data differences), it appears to compensate for it by inducing incorrect variations in the planetary spectrum, indicative of false opacity signatures and incorrect terminator temperature estimates for the atmosphere. This is also evident from the posterior probability distributions on the right panel of figure \ref{starcompareplot}, where the irradiation temperature, metallicity and carbon-to-oxygen ratio are significantly biased from the truth values of the planetary atmosphere.

With this important result, we demonstrate that while we can correct for stellar heterogeneity by simply parameterizing for the contamination spectrum and marginalizing over them within a retrieval, additional uncertainties must be accounted for based on our confidence in the accuracy of stellar atmosphere models themselves.

\section{DISCUSSIONS $\&$ CONCLUSIONS}\label{sec:discandconc}
We have shown that stellar spot/faculae contamination will strongly influence our ability to retrieve atmospheric properties from JWST transmission spectra of sub-Neptune like planets orbiting M-dwarfs. We find that for a range of plausible near-infrared spectrophotometric precisions, there is significant bias when marginalizing over planetary parameters from the transmission spectrum if not correcting for these contamination sources appropriately. The bias appears to be less significant with decreased transmission feature signal-to-noise (e.g., muting due to clouds/metallicity). We also demonstrate that our ability to adequately marginalize over the stellar contamination parameters is only as good as the degree to which stellar photosphere models match reality.  

Within the context of our work, there is very little knowledge of physically reasonable coverage fractions of stellar heterogeneities as well as their distributions and occurance frequencies throughout the photosphere of an M dwarf in current literature. Priors for spot covering fractions we used were derived from I-band variability of 1$\%$ \citep{newton2016,rackham2018} or an order of 0.05 or less from variability in G and R photometric bands \citep{rockenfeller2006a,rockenfeller2006b}. Spectroscopic methods to measure TiO lines emerging from spot regions of M-dwarfs have produced dramatically different estimates for spot covering fractions when compared to methods such as Doppler Imaging, varying from $\sim$50$\%$ to $\sim$10$\%$ \citep{oneal1998,barnes2011} respectively. A significant source of uncertainty with photometric variability measurements is also the assumption that starspot patterns vary as a function of rotation \citep{browning2010,barnes2011}, however there is only a weak correlation between rotation and activity for M-dwarfs according to \citet{reiners2010}. Moreover, uncertainties associated with the frequency of the occurrence of homogeneously scattered stellar spots provides difficulties in performing multi-wavelength stitching of non-simultaneous observations (e.g. \citealt{barstow2015stitch} and \citealt{bruno2018}) as we could potentially expect changes in the stellar contamination spectra in a time-dependent manner. 

Several uncertainties arise to determine accurate temperature contrast ratios between stellar photospheric and heterogeneity regions, as these have previously only been modeled assuming single spots and not for uniformly covered spot regions \citep{barnes2011,afram2015}. However, \citealt{barnes2011} show a clear relation between stellar activity induced jitter in RV observations as a function of wavelength--especially at low temperature contrasts between the stellar photosphere and spot regions. Leveraging this relation could potentially provide one avenue for acquiring observational constraints on temperature contrasts between the photosphere and heterogeneity regions of M-dwarfs. 

Numerous methods exist to estimate M-dwarf photospheric inhomogeneaities, however it still remains a challenging problem. The aim of this work is not to debate the presence of stellar contamination but to approach the problem from a ``marginalization-in-retrievals" perspective. By doing so we found that:

\begin{enumerate}
    \item \textit{There are degeneracies between photospheric contamination properties and transiting planet atmosphere properties within the transmission spectra:}
    
    In section \ref{sec:analytical} we gained a qualitative intuition for the expected degeneracies. From the relations in section \ref{sec:analytical} it is apparent that  stellar contamination $\epsilon_\lambda$ and planetary scale height $H$ are degenerate in that both result in a ``stretching'' effect in the spectrum at longer wavelengths. Ignoring stellar contamination therefore will lead to incorrect estimates of the inferred terminator temperature or molecular weight of the planetary atmosphere. At shorter wavelengths in particular, additional degeneracies emerge--between the stellar contamination temperature contrast C$_s$, covering fraction in area f$_s$ and the wavelength-dependent radius $z_\lambda$. Therefore, stellar contamination parameters can alter the apparent opacity structure of the planetary spectrum in a wavelength-dependent manner.
    
    \item \textit{There can be significant biases in the near-infrared transmission spectra for sub-Neptune planets orbiting M-dwarfs for a range of stellar contamination scenarios:}
    
    We showed that biases can exist in the atmospheric properties of interest, here, the metallicity, C-to-O ratio, and scale height temperature (see figures \ref{histoscenarioplot} and \ref{clear_pairs_plot}).  Failure to account for contamination within a retrieval would result in incorrect atmospheric parameter estimates, by several sigma, depending on the degree of the heterogeneities and the planetary feature signal-to-noise (e.g., muted vs. non-muted atmospheres).  We also showed through model comparison (the Bayes factor) that the evidence favoring the inclusion of a contamination parameterization increased with increasing spectro-photometric precision  and contamination fractions larger than 1$\%$. We also identified over the 2.5 - 5 $\mu$m spectral window notable degeneracies between the spot covering fraction and the atmospheric C-to-O ratio as well and gray cloud opacity, and also a degeneracy between the stellar spot temperature and the atmospheric metallicity.  

    \item \textit{There is no substantial improvement in the atmospheric parameter constraints if we have {\it a-priori} knowledge of the spot contamination characteristics :}
    
    Assuming the spot parameterization and stellar models are accurate representations of the true stellar photosphere, simultaneously retrieving the contamination parameters from the stellar spectrum directly (e.g., the Joint-fit model) does not provide a statistically significant improvement in the precision on atmospheric parameters. Simply including stellar contamination parameterization as a set of nuisance parameters within the transmission spectrum retrieval alone is sufficient. 
    
    \item \textit{An incorrect parameterization/representation of the contaminated stellar-photosphere can result in a substantial bias on the atmospheric properties:}
    
    Mimicking the differences between model and measured stellar spectra, we find that despite including corrections for stellar contamination (see figure \ref{starcompareplot}), retrievals would result in significantly biased atmospheric properties. This is evident for the regime that is sensitive to detect bias within the retrieval framework for planetary parameters i.e. f$_s$ $>$ 1$\%$. Therefore, our ability to correct for stellar contamination on the transmission spectrum is only as good as the accuracy of the stellar models. Understanding gaps within the stellar atmosphere models would be an important step towards alleviating this issue.

 Characterization of exoplanet atmospheres is a rewarding but challenging venture. Upcoming precision facilities/instrumentation will greatly improve our understanding of exoplanet atmospheres, but will also bring numerous new challenges. This work has only focused on the impact of a single ``degeneracy" on our ability to infer atmospheric properties  (namely the transit light source effect, e.g. \citealt{apai2018,rackham2018}). This is but one in a large realm of possible retrieval degeneracies/complications of varying degrees that include, but are not limited to, the spectral resolution linked bias \citep{deming2017}, treatment of opacities \citep{baudino2017}, cloud model assumptions (e.g., \citealt{mai2019,helling2008consistent,helling2008dust,helling2016mineral}), choice of abundance parameterizations \citep{kreidberg2015,kirkpatrick2016}, and three-dimensional effects \citep{LineParm2016,Feng2016,blecic2017}. Diagnosing the impact and improving of our understanding of these biases/degeneracies is critical to our determination of the true nature extra-solar planetary atmospheres.       
 
\end{enumerate}

\acknowledgments
We thank our referee for their extremely thorough and timely feedback on our manuscript.  We thank Jennifer Patience, Patrick Young, Evgenya Shkolnik, Mark Swain, Jeff Valenti, and Nestor Espinoza for meaningful discussions about this project. MRL \& ARI acknowledge support from NASA Exoplanet Research Program award NNX17AB56G and HST-GO-14793 and 15255. ARI also acknowledges support from the ASU SESE Summer Exploration Graduate Research Fellowship. The authors acknowledge Research Computing at Arizona State University for providing HPC \& storage resources that have contributed to the research results reported within this paper. URL: http://www.researchcomputing.asu.edu. This research has made use of the Exoplanet Orbit Database, the Exoplanet Data Explorer at exoplanets.org \citep{exoplanet.org} and the pysynphot database, URL https://pysynphot.readthedocs.io/en/latest/”. ARI would also like to thank Edward Buie II and Sean Czarnecki for proofreading the manuscript. 

\textit{Supplementary information}: zenodo link with all posterior probability distributions of all simulation scenarios discussed in this paper will be made available upon publishing.

\facilities{HPC Agave cluster at Arizona State University}
\software{PyMultiNest (Buchner et al. 2014)}

\bibliography{biblio}
\bibliographystyle{aasjournal}



\end{document}